# Clumpy Cold Dark Matter and Biological Extinctions

J. I. Collar

Department of Physics and Astronomy
University of South Carolina
Columbia, South Carolina 29208, USA

and

Groupe de Physique des Solides (UA 17 CNRS)
Universités Paris VI / VII
2, Place Jussieu, 75251 Paris Cedex 05, FRANCE

e-mail: ji.collar@sc.edu

**Abstract:**
**Cosmological models with cosmic string and texture seeded universes predict a present abundance of very dense clumps of Cold Dark Matter particles. Their crossing through the solar system would induce a non-negligible amount of radiation damage to all living tissue; the severity of such an episode is assessed. The estimated frequency of these crossings agrees with the apparent periodicity of the paleontological record of biological extinctions.**

It has been recently argued that clumpiness is a generic characteristic of Cold Dark Matter (CDM) cosmologies [1,2]. Silk and Stebbins [1] have analyzed the formation and posterior evolution of dark matter clumps in the cosmic string [3], texture [4] and inflationary models. The perturbation created around a moving nonlinear seed creates CDM overdensities, i.e., CDM clump cores. Secondary infall of CDM onto these cores cloaks them with virialized halos of smaller density. These halos are tidally stripped as clumps accrete onto larger and larger structures. This tidally stripped dark matter finds itself smoothly distributed and forming the galactic dark halos [1]. The average dark matter density in the laboratory would then be close to the present estimate ($\rho_{local} \sim 0.4$ GeV / c$^2$ cm$^3$) most of the time [1] and the surviving compact clump cores would roam the galaxy. These cores have radii $R_c \sim 0.001$ pc $m_\chi^{-1/2} \Omega_0^{-1} h^{-2}$, where $m_\chi$ is the mass of the CDM particle in GeV / c$^2$, $\Omega_0$ is the present-day density parameter and $h$ is the normalized Hubble parameter. The dependence on temperature at decoupling and relativistic degrees of freedom is dropped for simplicity, adopting their fiducial values [1]. The cores have masses $M_c \sim 0.02\ m_\chi^{-3/2} \Omega_0^{-1} h^{-2}$ (solar mass units) and the velocity dispersion of the CDM particles inside the core is a mere $v_c \sim 0.2\ m_\chi^{-1/2}$ km / s. The resulting core density is $\rho_c \sim 4.5 \cdot 10^8\ \rho_{local}\ \Omega_0^2\ h^4$. This density is estimated to be larger by a factor $\sim 10^2$ in ref. [2]; the above value is conservatively adopted in this discussion. Aside from stars, planets and molecular cloud cores, these surviving clump cores would be the densest objects in the galaxy [1].

The period associated with CDM core crossing through a given point in the galaxy can be estimated as $\tau = (\Phi \sigma)^{-1}$, where $\Phi$ is their local flux and $\sigma$ is the core's geometrical cross section. It is reasoned [1] that in the string-seeded case the cores cannot be drawn into the galactic center by dynamical friction and that their presence should be rather inconspicuous. In absence of a more exhaustive model one can assume that their spatial distribution follows the density profile of the galactic halo and that their velocity distribution is a broad Maxwellian with a dispersion velocity $v_{disp} = <v> \approx 300$ km / s, typical of a galactic collisionless gas. With this, and defining the fraction of all galactic CDM still in clump cores as $f$, one obtains the period of their crossing at our local galactic position as $\tau = [(f \rho_{local} / M_c)\ v_{disp}\ \sigma]^{-1} \sim 1.96 \cdot 10^9$ yr $m_\chi^{-1/2} f^{-1} \Omega_0 h^2$ (fig. 1). At least in the string scenario, where the original mass of a clump's halo is $\lesssim 10$ times that of its core [1] (suggesting $f \gtrsim 0.1$ in the galactic halo formation process described above), it is immediately striking that for $m_\chi \sim 10^2 - 10^4$ GeV / c$^2$ this period is similar to those found in statistical analyses of the paleontological record of biological extinctions, i.e., $\tau \sim 30 - 100$ Myr [5,6]. The average duration of core crossings is (using $R_c$ and $v_{disp}$) $t \sim 3.26$ yr $m_\chi^{-1/2} \Omega_0^{-1} h^{-2}$ (fig. 1).

Applying the assumed Maxwellian velocity distribution (similar to that of diffuse CDM particles in the galactic halo) it is possible to calculate the mean number of nuclear recoils per kg of organic tissue per day induced by a passing clump core composed of Weakly-Interacting Massive Particles (WIMPs). This is accomplished by substitution of $\rho_{local}$ by $\rho_c$ in the customary formalism of experimental searches for WIMP direct detection [7], also yielding the average nuclear recoil energy, $<T>$. The average dose rate imparted is obtained as the product of this scattering rate and $<T>$ (fig. 1). $C_4H_{40}O_{17}N_1$ is used as a representative tissue composition and $v_c$ ($<< v_{disp}$) is neglected. Heavy "neutrinos" (i.e., a generic massive neutral particle [7]) have been used as the WIMP in the calculation. The results apply to neutralinos [8] with predominantly scalar (spin-independent) couplings, for which the collision kinematics are the same (the dose rate must be scaled by their scattering total cross sections).

There is recent and mounting evidence that high-Linear Energy Transfer (LET) radiations such as alpha particles, fast neutrons (via their nuclear recoils) and heavy ions are responsible for unique biological effects [9-11]. Not all types of radiation are equally effective in producing important damage at the chromosomal level. Of special importance are those radiation insults that create irreparable genetic damage without inactivating the cell, which then can become the "founder" cell of an aggregation of mutant or cancerous cells. Low-LET radiations such as gamma and X-rays disperse their energy over longer distances than their high-LET counterparts, which have characteristic densely-ionizing tracks. There is now a wide consensus that the critical property of radiations at low doses is determined by this spatial pattern of energy deposition over dimensions similar to those of DNA structures (few nm) [10]. For instance, the primary molecular damage of importance in mammalian cells is produced by a localized cluster of atomic interactions overlapping the DNA and producing a multiply damaged chromosomal site [11]. Track structure analysis of high-LET radiations shows energy concentrations at subcellular levels that cannot be at all reproduced by low-LET radiation. Energy depositions of ~ 800 eV within 5 - 10 nm are unique to alpha particles and heavier ions, totally unattainable for other types of radiation and provoking unrepairable genetic damage [11]. In this respect, a damage irreparability threshold for mammal cells under ion irradiation has been established at LET > 100 keV / µm and a maximal neoplastic cell transformation rate (malignant cell transformation, the first step in tumour formation) has been found at LET ~ 100 - 200 keV / μm [12]. More recently, a striking transmission of non-clonal radiation-induced aberrations to clonal descendants of bone marrow cells irradiated by alpha particles (121 keV / μm) has been observed [13] and independently confirmed [14]. The lesions were considered able to produce the onset of leukaemias [13] and the chromosome imbalances and rearrangements to be like those observed in solid tumours [14]. Sister chromatid exchanges have been produced with very high effectiveness by similar high-LET irradiations in human lymphocytes [15].

Fig. 2 displays $<T>$ for the oxygen recoils produced by clump crossing and their corresponding LET in tissue (as obtained from the computer code TRIM92 [16]). This LET is very close to the maximum Relative Biological Effectiveness (RBE) for malignant cell transformation at ~ 100 - 200 keV / μm [12]. Oxygen constitutes ~ 90 % of all WIMP recoils, regardless of the value of $m_\chi$, due to the coherent scattering cross section (proportional to the square of the number of neutrons in the target nucleus [7]).

In order to establish a reference frame for the biological damage caused by a core crossing, we recall that the fast neutron component of the secondary cosmic-ray spectrum is able to induce similar recoils, albeit not uniformly distributed as in the WIMP case, comparatively diminishing their effect on highly radiosensitive tissue such as lymphatic cells or bone marrow. The measured cosmic neutron sea-level dose rate is ~ $10^{-6}$ Gy / yr [17], much smaller than the core crossing dose rate (fig.1). The ambient response of bubble neutron detectors [18], devices sensitive only to high-LET radiations, indicates that cosmic-neutron recoils account for most of the high-LET radiation dose in the terrestrial environment. Fast neutron recoils are also qualitatively different from WIMP recoils: more than 90% involve hydrogen atoms, with LET values well below maximum RBE and the irreparability threshold, and an energy loss mostly due to ionization. WIMP recoils, by contrast, lose an important fraction of their energy via atomic collisions (fig. 2), which are arguably more effective in creating DNA disruption [19,20]. Another comparison can be established with respect to fast neutrons from a nuclear detonation, generally responsible for most of the late radiation effects [21]: the instantaneous fast neutron dose to internal organs at 1.5 km ground range in Hiroshima was ~ $10^{-3}$ Gy [22]. In view of the essential differences between WIMP and neutron recoils, one can in principle naively relate the effects of a WIMP core crossing to those of neutron radiation from a close nuclear explosion, affecting all living creatures and protracted over $t$. Dose protraction may constitute another aggravating feature [10,23]. Further study into the biological effects of high-LET radiation will appraise the fitness of this comparison.

Even for the conservative value of $\rho_c$ adopted here, there is no question that the passing of a WIMP clump core would induce a large dose of highly mutagenic radiation to all living tissue, maybe explaining the observed bursts in diversification of life after extinctions [6]. The largest uncertainty is in the periodicity of these crossings, due to its dependence on $f$. Improved constrains on this fraction of galactic CDM still in clump cores will determine the relevance of the discussed process in explaining mass extinctions. Searches for geological evidence of a largely increased CDM flux coincidental with extinction periods would then be in order; nuclear tracks from WIMP recoils in ancient mica crystals is a possibility [24], but tracks from WIMP annihilation products such as proton-antiproton pairs [25] might leave a clearer signature against any constant background accumulated over the crystal's age. The gamma ray flux from WIMP annihilation in clumps has been suggested as a tool to constrain $f$ [1,2]. A possible characteristic contribution of CDM cores to gravitational microlensing or picolensing [26] might also limit or define $f$.

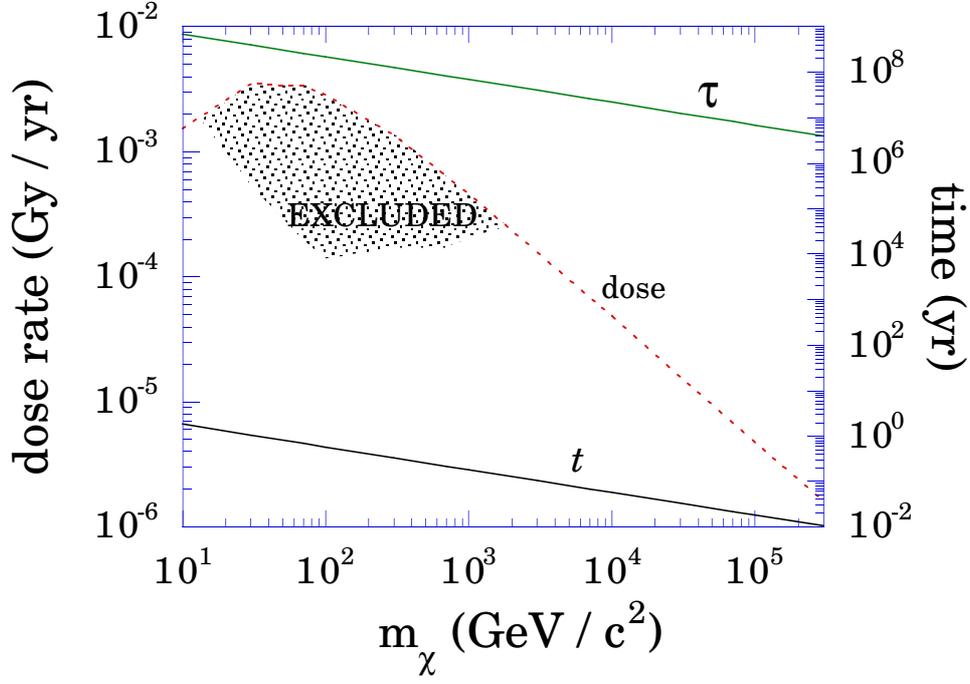

Fig. 1      Average dose rate in tissue ($\sim C_4H_{40}O_{17}N_1$) imparted via nuclear elastic scattering of the CDM particles constituting a clump core (1 Gy = $6.24 \cdot 10^{12}$ MeV / kg). Heavy Dirac neutrinos are used as the CDM particle [7]. The results can be scaled to neutralinos with dominant scalar couplings (see text). The shaded region corresponds to particle masses and couplings already excluded by underground germanium experiments [7]. Note that these exclusions are relaxed as $f$ approaches unity. The relevant parameter values used are $h = 0.75$, $\Omega_0 = 1$, $f = 0.5$. The period $\tau$ of core crossings through the solar system and their average duration $t$, are also shown.

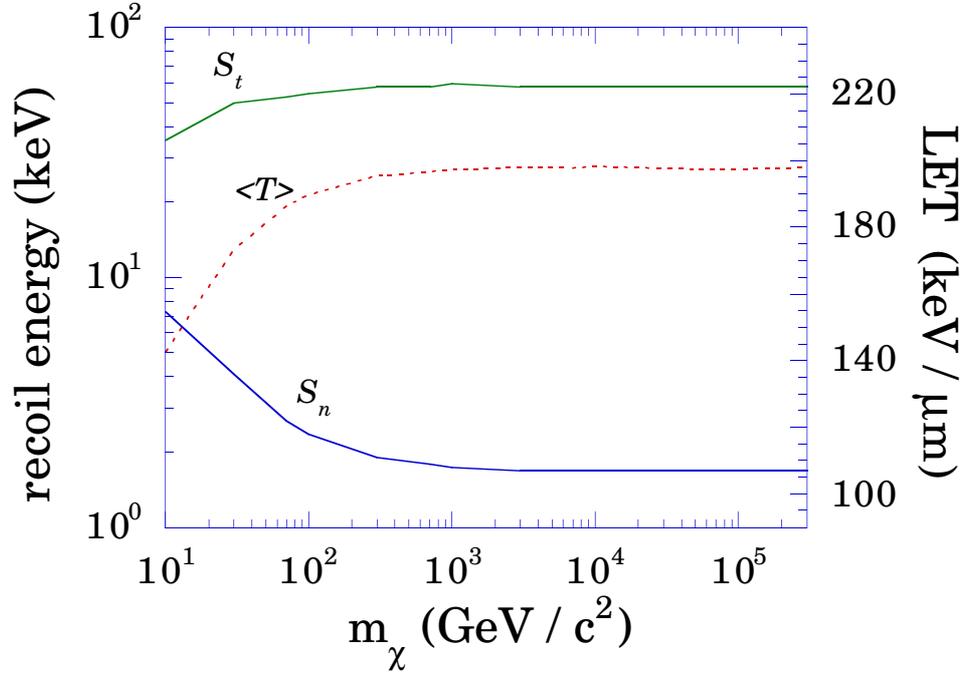

Fig. 2      Average recoil energy, $<T>$, of the predominant CDM oxygen recoils and their corresponding total stopping power, $S_t$, in tissue (equivalent to the unrestricted LET, the amount of energy dissipated by a radiation per unit path length). The component of the stopping power due to direct atomic collisions (as opposed to ionization), $S_n$, is also depicted.